\documentclass[aps]{revtex4}

\usepackage{graphics}
\usepackage{graphicx}
\usepackage{amsmath}
\usepackage{amsfonts}
\usepackage{mathbbol}
\newcommand{\szero}{s_{\scriptscriptstyle 0}}
\newcommand{\Dbn}{{\cal D}_{B}^{(n)}}
\newcommand{\hDbn}{\hat{\cal D}_{B}^{(n)}}

\begin{document}

\title{Effective theories of scattering with an attractive inverse-square
potential\\ and the three-body problem}

\author{Thomas Barford}
\author{Michael C. Birse}

\affiliation{Theoretical Physics Group, Department of Physics and
Astronomy\\
University of Manchester, Manchester, M13 9PL, UK\\}
\date{\today}


\begin{abstract}
A distorted-wave version of the renormalisation group is applied to
scattering by an inverse-square potential and to three-body systems.
In attractive three-body systems, the short-distance wave function satisfies
a Schr\"odinger equation with an attractive inverse-square potential, as
shown by Efimov. The resulting oscillatory behaviour controls the
renormalisation of the three-body interactions, with the 
renormalisation-group
flow tending to a limit cycle as the cut-off is lowered. The approach
used here leads to single-valued potentials with discontinuities as
the bound states are cut off. The perturbations around the cycle start
with a marginal term whose effect is simply to change the phase of the
short-distance oscillations, or the self-adjoint extension of the singular
Hamiltonian. The full power counting in terms of the energy and two-body
scattering length is constructed for short-range three-body forces.

\end{abstract}

\maketitle

\section{Introduction}

The successful application of effective field theories (EFT's) to two-body
scattering has revived interest in developing model-independent treatments
of few-body systems, as reviewed in Refs.~\cite{border,bvkrev}. In the case
of two-body systems EFT's provide a framework within which the old idea
of the effective-range expansion \cite{bethe,bj} can be extended 
systematically
to describe electromagnetic or weak couplings. Their application to 
three-body
systems requires the addition of three-body forces. The observation that
such forces are essential for describing low-energy three-body observables
goes back to the work of Phillips \cite{phillips}. One model-independent way
of introducing them is the boundary-condition method developed by Brayshaw
\cite{brayshaw}. The EFT approach provides a more practical way of doing so,
and one which can be extended to include couplings to external currents.

In the case of two-body scattering, these EFT's have been well explored.
If there is a clear separation between the low-energy scales of interest and
the high-energy scales characterising the underlying physics, then the
interactions terms in the effective Lagrangian can be organised 
systematically
as an expansion in powers of ratios of low-energy to high-energy scales.
For weakly interacting systems the power counting in this expansion is just
that of naive dimensional analysis. Since this is the same as the counting
in chiral perturbation theory \cite{wein79,wein}, it is usually known as
``Weinberg counting''.

In contrast, in strongly interacting systems with shallow resonances or
bound states, simple dimensional analysis is no longer appropriate because 
of
the appearance of new low-energy scales. These result in a need to resum 
certain
terms \cite{wein,uvk}. This has been done within various frameworks
\cite{uvk,ksw,bmr,geg}, all of which lead to the the same power counting.
The resulting scheme is often referred to as ``KSW counting''.
For systems with short-range forces only, it is in fact equivalent to the
effective-range expansion developed by Bethe and others \cite{bethe,bj}.
This expansion is the relevant one for few-nucleon systems at low energies,
where the deuteron and the $^1S_0$ virtual state lie very close to 
threshold.
It may also apply to atomic systems where Feshbach resonances can be tuned
to give large scattering lengths.

Both of these power-counting schemes can be understood using the
Wilsonian renormalisation group (RG) \cite{wilson}. Fixed points of the RG
describe scale-free systems. The RG flow near these points can then be used
to define a power counting for organising the perturbations around them.
Weinberg counting arises from the expansion around the trivial fixed point 
of
the RG, corresponding to a noninteracting system. KSW counting, or the
effective-range expansion, is associated with a second, nontrivial fixed 
point
corresponding to a system with a bound-state exactly at threshold.

The RG approach can also be extended to systems with known long-range
interactions, such as the Coulomb or one-pion-exchange potentials, provided 
one
has identified all the low-energy scales. In Ref.~\cite{bb1} we applied this
to examples including Coulomb and repulsive inverse-square potentials, for 
which
well-defined distorted waves (DW's) exist. We refer to this extended
version as the distorted-wave RG (DWRG). In it, a cut-off is applied to the 
basis
of DW's of the known long-range potential so that that potential is 
unaffected by
the cut-off. The advantage of the method is that it provides a clean 
separation
between the short- and long-range physics. The resulting nontrivial fixed 
point
corresponds to a DW or ``modified'' version of the effective-range expansion
\cite{bethe,vhk}. In this paper, we use the DWRG to determine the
power-counting for three-body forces. A very brief account of these ideas 
was
previously presented in Ref.~\cite{bb2} and a much more extensive one can be
found in the Ph.D.~thesis, Ref.~\cite{tbthesis}.

The EFT description of systems of three particles must include three-body 
forces,
represented by six-point interactions in the Lagrangian. To develop a
model-independent treatment of these systems, we use a two-body EFT with
couplings fixed by two-body data, and augment it with three-body terms.
However a consistent power counting needs to be determined for these new
interactions.

Complications arise in three-body systems because there can be long-range
forces resulting from the exchange of one of the three particles. The range 
of
these forces is controlled by the two-body scattering lengths.
In weakly interacting systems, where the size of the scattering lengths is
determined by the scale of the underlying physics, these forces can be 
treated
as  short-range and Weinberg counting applies to the three-body 
interactions.
On the other hand, systems with shallow bound or virtual states have large
two-body scattering lengths and hence long-range forces. Hence for systems
where the two-body forces can be organised with KSW counting, the 
correspnding
power-counting for the three-body forces is difficult to determine.

Bedaque {\it et al.}~\cite{bvk,bhvk} were the first to look at the problem 
of
short-range  three-body forces in EFT's. Their method is based on the 
equation
originally derived by Skorniakov and Ter-Martirosian (STM) \cite{stm} for
systems with zero-range two-body forces. In higher partial waves and in
$s$-wave nuclear systems with spin or isospin $\frac{3}{2}$ they found that 
the
three-body interactions are irrelevant (in the technical RG sense that they
vanish as the cut-off is lowered). This is because centrifugal forces or the
Pauli exclusion principle act to ensure that these systems are insensitive
to short-range physics. In contrast, for three bosons in $s$-waves or three
$s$-wave nucleons with total angular momentum and isospin $\frac{1}{2}$, 
there is
nothing to prevent all three particles from coming together. Bedaque
{\it et al.}\ found that three-body forces are not merely important in these
systems, they are in fact essential for producing well-defined results
\cite{bvk,bhvk} (see also Refs.~\cite{bbh,hm}).
More recently, a power-counting scheme for the three-body forces
in these systems has been constructed by renormalising the STM equation
order-by-order in the energy \cite{bghr}. Our work confirms this power 
counting
within the framework of a full RG analysis.

In other recent work on this problem, Phillips and Afnan \cite{pa} have
looked at the STM equation in the case of three $s$-wave bosons. They were 
able
to reproduce the results  of Bedaque {\it et al.}\ by using a subtractive
renormalisation and introducing a single piece of three-body data, namely 
the
three-body scattering length. The introduction of this piece of
data is equivalent to the LO three-body force of Ref.~\cite{bhvk}.
Similar results have also been obtained by Mohr \cite{mohr}.

An alternative to the STM equation for three-body systems with contact
interations is provided by the work of Efimov \cite{efimov} (for a review,
see Ref.~\cite{acp}). In general, Efimov's equations are difficult to solve
(see, for example, \cite{fedorov}) because they involve a nonseparable 
boundary
condition. However, in the case of infinite two-body scattering length, the 
equations
become separable in suitable cooordinates and can be solved exactly.
In this limit the Hamiltonian becomes scale free, which should not be 
surprising
given that this is the nontrivial fixed point of the two-body system.
In hyperspherical coordinates, the corresponding three-body Hamiltonian 
contains
an inverse-square potential (ISP).

The strength of the ISP is determined by the statistics of the system. In 
systems
with nonzero orbital angular momentum or where the Pauli exclusion principle
keeps the particles apart, the potential is repulsive. However if there is 
no
angular momentum or exclusion principle, as for three $s$-wave bosons, there
can be an attractive ISP. This corresponds to the cases where Bedaque
{\it et al.}\ found a need for a three-body force.

Because of its relevance to three-body EFTs, the attractive ISP has been the
subject of several recent papers \cite{bc,co,bp}. It will also play
a central role in this paper, since the DWRG analysis for scattering in the 
presence
of this potential will also provide the basic power counting for the 
three-body
system. The attractive ISP is particularly interesting from a mathematical 
point
of view because it is sufficiently singular that its wave functions have a
logarithmic oscillatory behaviour near the origin \cite{newton}. This makes 
it
impossible to define a ``regular'' boundary condition on the wavefunctions 
at the
origin. Mathematically, the resolution of this quandary is well known: we 
need to
form a self-adjoint extension of the Hamiltonian \cite{meetz,case,pp}. More
physically, one can think of this as introducing a boundary condition to fix
the phase of the oscillatory solutions near the origin.

Although choosing a self-adjoint extension ensures that all observables are
uniquely defined, it does not lead to a ground state. Instead the bound 
states
form an infinite tower with geometrically spaced energies. This pattern was 
found
by Efimov in the context of three-body systems \cite{efimov}, although the 
lack of
a ground state was noted first by Thomas \cite{thomas}. The pattern is not 
quite
scale-free since it requires the input of a new scale, such as the energy of 
one
of the bound states. This scale is provided by the self-adjoint extension.

An alternative to introducing a boundary condition to fix the self-adjoint
extension of the ISP is to regulate the singularity of the potential at 
short
distances and introduce a counterterm \cite{beane,bc}. In essence this is
what the momentum-space regularisation of the STM equation does for the 
three-body
problem \cite{bhvk,hm,bghr}. Again, one piece of three-body data is needed 
to fix
the scale in the extension, as also noted by Phillips and Afnan \cite{pa}.
Other authors have shown explicitly how a short-range force can be used to
select an extension \cite{bc,beane,bp}, but have not developed the power 
counting
for all possible terms in that force. Furthermore, the three-body forces 
obtained
in this manner turn out to be multi-valued, with no obvious procedure to 
resolve
this \cite{bc,bp}.

An interesting feature of the approaches based on regularisation of the 
singular
interaction is that they exhibit a cyclic behaviour as the scale of the
cut-off is varied \cite{bhvk,hm,beane,bc}. In fact, as pointed out by Wilson
\cite{wilslc}, they form a novel kind of limit cycle of the RG. Several
examples of this have only recently been highlighted \cite{gw,leclair,mh}.
In addition, Braaten and Hammer have pointed out that a rather minor fine 
tuning
of the quark masses in QCD would bring the infrared limit of the theory to 
the
effective-range fixed point in the two-nucleon channel, and hence to a limit 
cycle
for three nucleons \cite{bh}.

In this paper we apply the DWRG method, first to the attractive ISP 
corresponding
to three-body systems with infinite two-body scattering length, and then to
more general attractive three-body systems. We apply a cut-off to the deeply
bound states as well as the continuum, and this leads to an RG equation with
single-valued but discontinuous solutions. As already mentioned, the results
confirm the power counting found by Bedaque {\it et al.}~\cite{bghr}. By 
taking
advantage of Efimov's separation of the full three-body wave function, we 
are
able to show that this power counting is not limited to the STM ``slice'' 
through
it, where two of the particles coincide.
Because of the clean separation of short- and long-range physics in the 
DWRG, we
are able to derive this counting in a more transparent way.

The lowest-order three-body interaction is marginal, in the sense that its 
RG
flow is only logarithmic in the cut-off. As we shall explicitly show, it
represents the same degree of freedom as the choice of self-adjoint 
extension.
The resulting three-body interaction is built out of perturbations around
a single-valued limit-cycle solution of the DWRG. By constructing the
corresponding physical scattering amplitudes, we are able to directly relate
the coefficients in the three-body potential to observables, through a
version of the DW effective-range expansion.

\section{The RG for three-body forces}

For simplicity, we consider here a system of three bosons of mass $M$, in a
state of zero orbital angular momentum. We assume that the two-body 
interaction
produces a single shallow bound state with binding energy $\gamma^2/M$,
corresponding to a ``binding momentum'' $i\gamma$.
We start by considering the effects of two-body forces only and we denote
the corresponding three-body Green's function by $G_2(p)$, where $p$ is 
defined
in terms of the total centre-of-mass energy by $E=p^2/M$.  This function
has the spectral decomposition,
\begin{equation}\label{eq:fullgreen}
G_2(p)=\frac{M}{4\pi}\sum_{n=1}^N\frac{|\Psi_n\rangle\langle 
\Psi_n|}{p^2+p_n^2}
+\frac{M}{2\pi^2}\int_{-\gamma^2}^{\infty}d(q^2)
\frac{1}{p^2-q^2+i\epsilon}\left[|\Psi_{q,i\gamma}\rangle\langle 
\Psi_{q,i\gamma}|
+\vartheta(q^2)\frac{2}{\pi}\int_0^q dk\,|\Psi_{q,k}\rangle\langle 
\Psi_{q,k}|\right],
\end{equation}
where $\vartheta(x)$ is the unit step function.
Note that in a system with two-body contact interactions, we require an 
additional
boundary condition to define the DW's and hence the Green's function, as 
discussed
below.

Three types of wave appear in this decomposition. In the first term, we have 
the
bound states of three particles, $|\Psi_n\rangle$. In the final term, we 
have
states with three incoming free particles, $|\Psi_{q,k}\rangle$,
which we label by their total centre-of-mass energy, $E=q^2/M$, and relative
momentum, $k$. In the middle are states with one incoming free particle and 
a
bound pair  $|\Psi_{q,i\gamma}\rangle$. These states are normalised as 
described
in Appendix A.

In this paper we work with the coordinate-space representation of the DW's.
Our EFT's are expressed in terms of contact interactions and so we are 
interested
in the region where all three particles are close together. A natural 
measure of
the proximity of the three particles is the hyperradius, $R$, defined in 
Appendix
A. We write the DW's as functions of $R$ and the five hyperspherical angles,
collectively denoted by $\Omega$, which represent the other degrees of 
freedom in
the centre-of-mass frame. The precise specification of these angles is
not needed since all $\Omega$ dependence will factor out of our results.

Now consider the effect of introducing a three-body interaction, $V_3$.
This could be inserted directly into the Faddeev equations \cite{m3hp}.
However, for our purposes, it is sufficient to use the ``two-potential 
trick''
\cite{newton} to define a $T$-matrix for the additional scattering produced
by $V_3$. We write a full $3\rightarrow3$ $T$-matrix, $T(p)$, in terms of 
the
one with two-body forces only, $T_2(p)$, plus an additional 
piece, $\tilde T_3(p)$, that acts between the DW's for the two-body forces:
\begin{equation}\label{eq:tdecomp}
\langle q,k|T(p)|q',k'\rangle=\langle q,k|T_2(p)|q',k'\rangle+
\langle\Psi_{q,k}|\tilde T_3(p)|\Psi_{q',k'}\rangle.
\end{equation}
The DW term $\tilde T_3(p)$ satisfies the Lippmann-Schwinger (LS) 
equation,
\begin{equation}\label{eq:tlse}
\tilde T_3(p)=V_3+V_3G_2(p)\tilde T_3(p).
\end{equation}
There is no problem with connectedness in this equation since the kernel 
contains
the three-body force.

For zero-range two-body interactions, the small-$R$ behaviour of the
wave functions in Eq.~(\ref{eq:fullgreen}) can be found from the approach of
Efimov \cite{efimov}. The essential elements of this, in our notation, are
outlined in Appendix A. Since the wave functions do not have well-defined 
limits
as $R\rightarrow 0$, we choose our effective three-body interaction to act 
at
some small, but nonzero hyperradius, $R=\bar R$. At this point the 
separations of
all three particles are less than or of the order of $\bar R$. This 
additional
regularisation is quite separate from the running cut-off which will give 
the RG
flow. As with the singular long-range potentials studied in Ref.~\cite{bb1}, 
it
is needed to make all wave functions appearing in the RG equation 
well-defined.
Provided we choose $\bar R$ to be in the region where the waves have reached
their asymptotic small-$R$ forms, the common dependence on $R$ can be 
factored
out of the problem. One could therefore imagine taking the $R\rightarrow 0$ 
limit
of the results, corresponding to a zero range force.
In addition, we take the interaction to act at some fixed set
of hyperangles $\Omega=\bar\Omega$. Since the wave functions at small $R$ 
are
separable, all hyperangular behaviour can be factored out and this arbitrary
choice will not affect the results. We can therefore define
\begin{equation}\label{eq:potform}
\langle\Psi_{p,k}|V_3|\Psi_{p,k'}\rangle=\bar R^4\Psi_{p,k}^*(\bar 
R,\bar\Omega)
\Psi_{p,k'}(\bar R,\bar\Omega)\,V_3(p,k,k').
\end{equation}

In an EFT we aim to integrate out all the unknown short-range physics and
replace it by a three-body force which can be expanded in powers of the 
low-energy
scales of the system. The tool which allows us to determine the scaling 
behaviour
of the terms in $V_3$ is the renormalisation group (RG) \cite{wilson}.
The first step in setting this up is to impose a cut-off, $\Lambda$,
to separate the low-energy states which we wish to treat explicitly
from the high-energy states which will be integrated out. We then 
renormalise
the theory by making the effective interaction $\Lambda$-dependent and 
demanding
that observables should be independent of the cut-off.

As in the approach developed in Ref.~\cite{bb1}, we impose this cut-off on 
the
DW's of our pairwise forces. By applying the cut-off to these waves rather 
than
the apparently simpler free waves, we ensure that the role of the three-body
force is related purely to short-range three-body physics. If the cut-off
were applied to the free waves, it would also remove parts of the physics
associated with the two-body forces. This would have to be compensated by 
the
three-body force, which would then satisfy a far more complicated
evolution equation.

We impose the renormalisation condition that amplitude $\tilde T_3$ be
independent of cut-off, $\partial_\Lambda\tilde T_3=0$. Differentiating
the LS equation (\ref{eq:tlse}) for $\tilde T_3$ then leads, after the
elimination of $\tilde T_3$, to a differential equation for $V_3$,
\begin{equation}
\frac{\partial}{\partial\Lambda}V_3(\Lambda)
=-V_3(\Lambda)\,\frac{\partial 
G_2(p,\Lambda)}{\partial\Lambda}\,V_3(\Lambda).
\label{eq:potdiff}
\end{equation}
The RG equation is then obtained by rescaling the potential $V_3$ and
and all the low-energy scales in this equation. As described below, the
definition of the rescaled potential $\hat V_3$ depends upon the form of the
DW's close to the origin. The boundary conditions on $\hat V_3$
are that it should been analytic in all rescaled energies and momenta.
These follow from our requirement that the terms in the three-body force
arise from local six-point vertex terms in the EFT Lagrangian.
The $\Lambda$-dependence of the corresponding terms in $\hat V_3$ then
allows us to classify them according to some power counting.

So far we have said nothing about the nature of the two-body interaction, 
except
that it leads to a single low-energy bound state. This could be represented 
as a
contact interaction, whose terms correspond to the effective-range 
expansion.
Equivalently it can be thought of as a boundary condition on the
logarithmic derivative of the wave function when two of the particles 
coincide.
Combining this with the overall symmetries of the wave function leads to
Efimov's boundary condition \cite{efimov}, as outlined in Appendix A.

This boundary condition becomes separable in the limit where the 
hyperradius, $R$,
is much smaller than the two-body scattering length, $a_2$. The Hamiltonian 
is
then diagonal in the hyperangular momentum variable, $s$. The potential in 
each
hyperangular-momentum channel is an ISP (as it must, since all scales have 
been
eliminated from the problem). The resulting Schr\"odinger equation can
be solved analytically in terms of Bessel functions of general order.

In the cases of three bosons in $s$-waves or two neutrons and a proton with 
total
spin $\frac{1}{2}$, the lowest hyperangular momentum gives rise to an 
attractive
ISP. As discussed in the introduction, there is an ambiguity in the wave 
functions
for this potential. In order to resolve this, we need to introduce
an additional boundary condition to form a self-adjoint extension.
As shown in appendix A, the form of the resulting DW's at small $R$ is
\begin{equation}
\Psi(R,\Omega)\sim\xi_{s_0}(\Omega)\sin\bigl(\szero\ln(p_*R)-\theta\bigr),
\label{eq:sepwfn}
\end{equation}
where $\szero\simeq 1.006$ is the magnitude of the smallest hyperangular 
momentum,
$\theta$ is defined in Eq.(\ref{eq:theta}), $\xi_{s_0}(\Omega)$ is the
hyperangular wavefunction, and $p_*$ is the scale introduced by forming a
self-adjoint extension. This form is invariant (up to an overall sign)
under the replacement,
\begin{equation}
p_*\rightarrow p_*e^{n\pi/\szero},\qquad n\in\mathbb Z.\label{eq:p*transf}
\end{equation}
As a consequence all observables are invariant under this transformation and 
so
all of the values of $p_*$ in Eq.~(\ref{eq:p*transf}) define the same
self-adjoint extension.

\section{Infinite scattering length}

In the limit of infinite two-body scattering length, the Hamiltonian due to
pairwise forces is separable. If we assume also that the three-body force is
diagonal in the hyperangular momentum then the RG equations for the various
channels are uncoupled. This approximation neglects parts of the interaction
that couple the channels. We introduce it here so that we can focus on the 
lowest
hyperangular-momentum channel, which contains the leading interactions.
Later, when we derive general results, we shall not require it.

The higher hyperangular-momentum channels, $s_i$ with $i>0$, correspond to
repulsive ISP's. Their RG analysis is thus identical to that described in
Ref.~\cite{bb1}. It leads to a power-counting scheme in which the LO term in
the force scales with $(p/\Lambda_0)^{2s_i}$, where $\Lambda_0$ is the scale 
of
the underlying physics. The most important of the interactions based upon 
this
power counting is the LO term in the $s_1=4$ channel, but even this term is
heavily suppressed. Similarly, in systems which do not lead to attractive 
ISP's,
the lowest-order three-body forces are always suppressed by powers of
$p/\Lambda_0$. Here we concentrate on the $\szero$ channel in attractive 
systems,
where the effects of three-body forces are enhanced.

\subsection{RG equation}

The hyperradial Green's function in the $\szero$ channel is given by
\begin{equation}
g_2(p;R,R')=\frac{M}{2\pi}\sum_{n=-\infty}^{\infty}\frac{u_{\szero}^{(n)}(R)
u_{\szero}^{(n)}(R')}{p^2+p_n^2}+
\frac{M}{\pi^2}\int_0^\infty dq\frac{u_{\szero}(q,R)u_{\szero}(q,R')}
{p^2-q^2+i\epsilon}.
\end{equation}
We use lower case to denote quantities in this channel, for example this
Green's function and the potential $v_3$,  to differentiate
them from those for the full problem. The DW's $u_{\szero}(p,R)$ are
simply solutions of the ISP Schr\"odinger equation (\ref{eq:2dse}).

The bound states $u_{\szero}^{(n)}(R)$ have energies which, from
Eq.~(\ref{eq:bsmom}), form a tower with geometric spacing,
\begin{equation}
E_n=-\frac{p_*^2}{M}e^{2n\pi/\szero}.
\end{equation}
These respect the symmetry of Eq.~(\ref{eq:p*transf}). They accumulate at 
zero
energy and extend deeper and deeper with no ground state. The shallow
states are referred to as Efimov states, although the lack of a ground state 
had
been noted much earlier. In 1935, Thomas \cite{thomas} pointed this out for
three-body systems bound by contact two-body forces.

Before we can write down the DWRG equation for the three-body force in this
channel we must first decide how to handle the bound states.
In the previous examples studied with the DWRG method \cite{bb1}, the 
cut-off
was applied purely to the high-energy continuum states. In those examples,
any bound states were shallow, with typical momenta much smaller than
the scale of the underlying physics, and so lay within the domain of the 
EFT.
Here, however, the lack of a ground state means that this is no
longer true. The deeply bound states are consequences of our use of
contact interactions to represent the two-body forces, and hence are
unphysical artefacts. In keeping in the EFT philosophy these states should 
be
truncated and their effects absorbed into the effective three-body force.
We chose to cut-off the Green's function by removing all states with energy
outside the range $-\Lambda^2/M\le E \le\Lambda^2/M$:
\begin{equation}
g_2(p;R,R')=\frac{M}{2\pi}\sum_{|p_n|<\Lambda}\frac{u_{\szero}^{(n)}(R)
u_{\szero}^{(n)}(R')}{p^2+p_n^2}+
\frac{M}{\pi^2}\int_0^\Lambda dq\frac{u_{\szero}(q,R)u_{\szero}(q,R')}
{p^2-q^2+i\epsilon}.\label{eq:isgreen}
\end{equation}
We shall see that truncation of bound states will allow us to get an RG
equation with single-valued solutions.

Using the separable form, Eq.~(\ref{eq:sepwfn}), of the wave functions, we
can project the differential equation for the potential (\ref{eq:potdiff})
onto the $\szero$ channel. We can then absorb the common angle-dependent
factor $|\xi_{s_0}(\bar\Omega)|^2$ into into the potential by defining
\begin{equation}
v_3(p,\Lambda)=|\xi_{s_0}(\bar\Omega)|^2V_3(p,\Lambda).
\end{equation}
Note that we have chosen to focus on potentials that depend only on energy
since, as discussed in Refs.~\cite{bmr,bb1}, these contain the leading
perturbations.

Projecting the differential equation (\ref{eq:potdiff}) onto this channel
and inserting Eq.~(\ref{eq:isgreen}) we obtain an equation for $v_3$,
\begin{equation}
\frac{\partial v_3(p,\Lambda)}{\partial\Lambda}
=\frac{M}{\pi^2}\left[\frac{|u_{\szero}(\Lambda,\bar R)|^2}{\Lambda^2-p^2}
-\frac{\pi}{2}\sum_{n=-\infty}^\infty
\frac{|u_{\szero}^{(n)}(\bar R)|^2}{p^2+p_n^2}\delta(\Lambda-p_n)
\right]v_3(p,\Lambda)^2.
\label{eq:v3def}
\end{equation}
The first term is produced by cutting off the continuum states, and is 
similar
to the expressions in Refs.~\cite{bmr,bb1}. The series of discontinuities
represented by $\delta$-functions results from the truncation of the bound
states.

Finally, to obtain an RG equation we need to rescale any low-energy scales
in the problem. In this case we have only one, the on-shell momentum $p$,
and we define $\hat p=p/\Lambda$. To form a dimensionless potential, we 
multiply
by the mass $M$. We also take the expressions for the wave functions near
the origin, Eqs.~(\ref{eq:isqdwr0}, \ref{eq:isqbsr0}), and absorb the
$\bar R$-dependence into the potential by defining
\begin{equation}
\hat v_3(\hat p,\Lambda)=\frac{M}{\szero\pi^2}\,
\sin^2\bigl(\szero\ln(p_*\bar R)-\theta\bigr)
\,v_3(\hat p\Lambda,\Lambda).
\end{equation}

The rescaled potential $\hat v_3$ satisfies the RG equation
\begin{equation}\label{eq:dwrg}
\Lambda\frac{\partial\hat v_3}{\partial\Lambda}
=\hat p\frac{\partial\hat v_3}{\partial\hat p}+\left[\frac{\sinh(\pi 
\szero)}
{\bigl[\cosh(\pi \szero)-\cos\bigl(2\szero\ln(\Lambda/p_*)\bigr)\bigr]
(1-\hat p^2)}
-\frac{\pi}{\szero}\sum_{n=-\infty}^{\infty}
\frac{1}{1+\hat p^2}\,\delta\bigl(\hat p_n(\Lambda)-1\bigr)\right]\hat 
v_3^2,
\end{equation}
where $\hat p_n(\Lambda)=p_n/\Lambda=p_* e^{n\pi/\szero}/\Lambda$. Note that
we write $\hat p_n(\Lambda)$ as a function of $\Lambda$ since we do not 
rescale
the scale from the extenstion, $p_*$, and so $\hat p_n$ varies with 
$\Lambda$.
As noted before \cite{bb1}, this type of equation is more conveniently 
rewritten
as a linear equation for $1/\hat v_3$,
\begin{equation}\label{eq:dwrg2}
\Lambda\frac{\partial}{\partial\Lambda}\left(\frac{1}{\hat v_3}\right)=
\hat p\frac{\partial}{\partial\hat p}\left(\frac{1}{\hat v_3}\right)-
\frac{\sinh(\pi \szero)}{\bigl[\cosh(\pi \szero)
-\cos\bigl(2\szero \ln(\Lambda/p_*)\bigr)\bigr](1-\hat p^2)}
+\frac{\pi}{\szero }\sum_{n=-\infty}^{\infty}
\frac{1}{1+\hat p^2}\,\delta\bigl(\hat p_n(\Lambda)-1\bigr).
\end{equation}

This RG equation is the same as that for an attractive ISP in two 
dimensions.
In fact an attractive ISP in any number of dimensions leads to an equation 
of
this form, because the different real power of $\bar R$ which appears in
the short-distance wave functions exactly compensates for the radial factor
in the Jacobian. All that differs is a numerical factor in the definition
of $\hat v_3$, associated with the angular integration.

\subsection{Solutions}

The physically acceptable short-range potentials are given by solutions to
the RG equation (\ref{eq:dwrg}) which satisfy the boundary condition of
analyticity in $\hat p^2$ for small $\hat p$. We look first for fixed 
points,
$\Lambda$-independent solutions of the equation. The power counting for 
terms
in the potential can be determined from the perturbations around a fixed 
point
that scale with definite powers of $\Lambda$ \cite{bmr,bb1}. These are
eigenfunctions of the linearised version of the RG equation.

An obvious fixed point is the trivial one, $\hat v_3=0$. The power
counting based on it can be found by substituting
\begin{equation}
\hat v_3=C\Lambda^\mu\phi(\hat p)
\end{equation}
into Eq.~(\ref{eq:dwrg}), and linearising to get the eigenvalue equation
\begin{equation}
\hat p\frac{d\phi}{d\hat p}=\mu\phi.
\end{equation}
This equation is easily solved, giving $\phi(\hat p)=\hat p^\mu$. Imposing
the boundary condition of analyticity in $\hat p^2$, leads to the
eigenvalues $\mu=0,2,4,\ldots$. Hence the general solution in the region of
this fixed point is
\begin{equation}
\hat v_3(\Lambda,\hat p)=\sum_{n=0}^\infty C_{2n}\Lambda^{2n}\hat p^{2n}.
\end{equation}
The leading term in this expansion is marginal, that is, it does not scale 
with
any power of $\Lambda$. We therefore expect to find logarithmic dependence 
on
$\Lambda$ associated with this perturbation. To resum these logarithms we 
need to
construct solutions of the full nonlinear RG equation.

In fact the presence of $\Lambda$-dependence on the right-hand side of
Eq.~(\ref{eq:dwrg}) means that no other fixed-point solution can be found.
However that dependence is only logarithmic and so we may look for slowly
evolving solutions which also depend logarithmically on $\Lambda$.
Perturbations about such a solution can still be used to construct a power
counting.

As in Ref.~\cite{bb1}, our starting point for constructing a nontrivial
solution to the RG is the basic loop integral, which in this case is
given by
\begin{equation}\label{eq:defineI1}
\hat I(\hat p,\Lambda)={\cal P}\int^1_0 d\hat q\,
\frac{\hat q\sinh(\pi \szero )}{\bigl[\cosh(\pi \szero )-\cos\bigl(2\szero
\ln(\Lambda\hat q/p_*)
\bigr)\bigr](\hat p^2-\hat q^2)}.
\end{equation}
By substituting this for $1/\hat v_3$ in Eq.~(\ref{eq:dwrg2}), it is
straightforward to verify that it satisfies the continuous version
of the equation (without the discontinuities of the final term).
However it is not yet an acceptable solution since, apart from missing
the bound state discontinuities, it is nonanalyitic for small $\hat p$.

The propagator pole at $\hat q=\hat p$ in the integrand approaches the 
endpoint
of the integral as $\hat p\rightarrow 0$. This leads to logarithmic 
dependence
of $\hat I$ on $\hat p$. To avoid this troublesome endpoint, and also to
introduce the required discontinuities, we need to find a similar integral
over some contour in the complex $\hat q$-plane avoiding the singular 
region.

The first step is to rewrite the integrand of $\hat I$ as
\begin{equation}
\frac{1}{2i}\frac{\hat q}{(\hat p^2-\hat q^2)}
\left[\cot\left(\szero \ln\frac{\Lambda\hat q}{p_*}
-\frac{i\pi \szero }{2}\right)-\cot\left(\szero \ln\frac{\Lambda\hat q}{p_*}
+\frac{i\pi \szero }{2}\right)\right],\label{eq:splitI}
\end{equation}
so that $I$ becomes
\begin{equation}
I(\hat p,\Lambda)={\cal P}\int_{-1}^1 d\hat q\,h(\hat q),
\label{eq:jdef}
\end{equation}
where
\begin{equation}
h(\hat q)=\frac{1}{2i}
\frac{\hat q}{(\hat p^2-\hat q^2)}\cot\left(\szero \ln\frac{\Lambda\hat 
q}{p_*}
-\frac{i\pi \szero }{2}\right).
\end{equation}

\begin{figure}
\begin{center}
\includegraphics[width=10cm,angle=0]{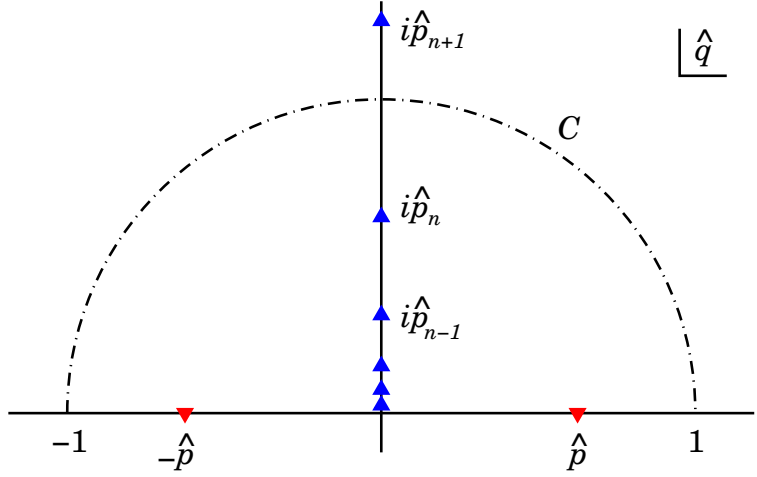}
\caption{The contour $C$ in the complex $\hat q$-plane used to construct the
solution $\hat\jmath(\hat p,\Lambda)$ of the DWRG equation. The bound-state 
poles
of the integrand occur at $\hat q=i\hat p_n =ip_*e^{n\pi/\szero }/\Lambda$
and the propagator poles at $\hat q=\pm\hat p$.}
\label{fig:jpoles}
\end{center}
\end{figure}

This integral has the same nonanalytic behaviour on the real axis as that in
Eq.~(\ref{eq:defineI1}). However, since it is no longer associated with an
endpoint of the integral, we are free to deform the contour of integration
into the complex plane to avoid the dangerous region around $\hat q=0$.
We therefore define
\begin{equation}
\hat\jmath(\hat p,\Lambda)=\int_C d\hat q\,h(\hat q),
\end{equation}
where $C$ is the contour of integration shown in Fig.~\ref{fig:jpoles}.
This runs from $\hat q=-1$ to $\hat q=1$ in the upper half plane and
crosses the imaginary axis at $\hat q=i$.

Fig.~\ref{fig:jpoles} also shows the pole structure of the integrand $h(\hat 
q)$.
Apart from the two propagator poles at $\hat q=\pm \hat p$, there are
bound-state poles at
\begin{equation}
\hat q=i\hat p_n(\Lambda)=i\,\frac{p_*}{\Lambda}\,e^{n\pi/\szero }.
\end{equation}
These poles are important since their positions vary with $\Lambda$ and so
they can cross our integration contour. When this happens they generate
discontinuities in $\hat\jmath(\hat p,\Lambda)$. By choosing our contour
to cross the imaginary axis at $\hat q=i$ (corresponding to an energy
$E=-\Lambda^2/M$) we can arrange for these discontinuities to match 
precisely
the ones we require in Eq.~(\ref{eq:dwrg2}). To see this, note that as 
$\Lambda$
is varied from $p_n-\epsilon$ to $p_n+\epsilon$, the pole at $\hat q=i\hat 
p_n$
crosses the contour and produces a discontinuity
\begin{equation}\label{eq:bsres}
\Bigl[\hat\jmath(\hat 
p,\Lambda)\Bigr]_{\Lambda=p_n-\epsilon}^{\Lambda=p_n+\epsilon}
=-2\pi i\,{\cal R}\left[h(\hat q),i\hat p_n\right]=\frac{\pi}{\szero }\,
\frac{1}{1+\hat p^2},
\end{equation}
where ${\cal R}[f(z),z_0]$ denotes the residue of $f(z)$ at the pole $z_0$.
This is precisely the strength of the $\delta$-function at $p_n(\Lambda)=1$
in Eq.~(\ref{eq:dwrg2}).

The integral $\hat\jmath(\hat p,\Lambda)$ satisfies the full RG equation
(\ref{eq:dwrg2}) and is analytic about $\hat p^2=0$. We therefore define the
rescaled potential by
\begin{equation}\label{eq:vs0}
\hat v_3^{(0)}(\hat p,\Lambda)=\bigl[\hat\jmath(\hat p,\Lambda)\bigr]^{-1}.
\end{equation}
This is not a fixed point since it does depend logarithmically on $\Lambda$,
but in it these logarithms have been resummed to all orders. The invariance 
of
the system under Eq.~(\ref{eq:p*transf}) means that this dependence on
$\ln\Lambda$ is periodic. As pointed out by Wilson \cite{wilslc}, this 
periodic
behaviour provides an example of a limit cycle of the RG.

Like a fixed point, the limit-cycle solution can be used to define
a power counting for the perturbations around it. Because the RG equation in 
the
form of Eq.~(\ref{eq:dwrg2}) is linear in $1/\hat v_3$, it is possible to
construct exact solutions containing these perturbations,
\begin{equation}\label{eq:solution}
\frac{1}{\hat v_3(\hat p,\Lambda)}=\frac{1}{\hat v_3^{(0)}(\hat p,\Lambda)}+
\sum_{n=0}^{\infty}C_{2n}\Lambda^{2n}\hat p^{2n}.
\end{equation}
The perturbations around the cycle thus have exactly the same power counting
as those around the trivial fixed point. The leading, energy-independent 
term
is marginal. Each additional power of energy ($\hat p^2$) gives a term two
orders higher in $\Lambda$.

The general solution, Eq.~(\ref{eq:solution}), shows that there is a family
of limit cycles parameterised by the marginal perturbation $C_0$. This lack 
of
uniqueness arises because the marginal perturbation is $\Lambda$-independent
and satisfies the homogeneous version of the RG equation (\ref{eq:dwrg2}).
As a result, an arbitrary amount of it can be added to the limit-cycle 
solution,
Eq.~(\ref{eq:vs0}). The leading perturbation around any cycle is marginal 
because
it pushes the solution into a nearby cycle. In the limit 
$C_0\rightarrow\infty$,
the entire cycle is compressed into the trivial fixed point $\hat v_3=0$, 
which
can be regarded as the limiting member of the family of cycles. All other
perturbations vanish as $\Lambda\rightarrow 0$ and hence are stable. Any
more general solution will thus tend to one of the limit cycles as
$\Lambda\rightarrow0$.

Each limit cycle contains one discontinuity every period, when a bound state 
is
removed from the low-energy domain. The point in the cycle where this occurs
depends on how we truncate the sum over bound states. Here we have chosen a
cut-off that is symmetric between positive and negative energies, although
other prescriptions are equally acceptable since physical observables should 
not
depend on them.

The truncation of bound states ensures that our solution of the RG equation 
is
single-valued. Our choice corresponds to a particular path
from $\hat q=-1$ to $+1$ for the contour $C$ used to evaluate
$\hat\jmath(\hat p,\Lambda)$. Other, topologically different paths would
correspond to different branches of a general multi-valued solution.
Such multi-valuedness appears in other methods used to renormalise the
three-body potential, as in Refs.~\cite{bhvk,bbh,hm,bghr}.

\subsection{Scattering observables}

To relate the parameters appearing in our solution, Eq.~(\ref{eq:solution}),
to scattering observables, we need to construct the corresponding
$T$-matrix. If we use the separable form of the wave functions, we can 
define a
DW $T$-matrix $\tilde t_3$ in an analogous manner to Eq.~(\ref{eq:v3def}).
Projecting the LS equation Eq.~(\ref{eq:tlse}) onto the $s_0$ channel,
we get the correspondimng equation for $\tilde t_3$. With our choice of
energy-dependent potential, this can be solved directly to get the on-shell
$T$-matrix,
\begin{equation}
\langle u^-_p|\tilde t_{3}(p)|u^+_p\rangle
=e^{2i\delta_2(p)}|u_{p}(\bar R)|^2 v_3(p,\Lambda)
\left[1-g_{2}(p,\Lambda;\bar R,\bar R)v_3(p,\Lambda)\right]^{-1},
\label{eq:ts01}
\end{equation}
where the superscripts $\pm$ denote waves with incoming or outgoing boundary
conditions, and $\delta_2$ is the phase shift in the $\szero$ channel 
produced
by the pairwise forces alone.

Unitarity and conservation of hyperangular momentum allow us to express this
in terms of a phase-shift as
\begin{equation}
\frac{e^{2i\delta_2(p)}}{\langle u^-_p|\tilde t_{3}(p)|u^+_p\rangle}
=-\frac{M}{2\pi p}
\bigl(\cot\tilde\delta_3(p)-i\bigr).\label{eq:ts02}
\end{equation}
Here $\tilde\delta_3$ is the additional phase shift produced by the 
three-body
force. It is related to the full phase shift $\delta$ by
$\tilde\delta_3=\delta-\delta_2$. Using the fact that our potential acts at
$R=\Bar R$, we can combine Eqs.(\ref{eq:ts01}, \ref{eq:ts02}) to obtain an
equation relating $v_3$ and the phase shift:
\begin{equation}
|u_{p}(\bar R)|^2\frac{M}{2\pi p}(\cot{\tilde\delta_3}-i)=
g_2(p,\Lambda;\bar R,\bar R)-v_3(p,\Lambda)^{-1}.
\end{equation}

When we use the expression in Eq.~(\ref{eq:isgreen}) for the regularised
Green's function and the explicit forms of the wave functions from Appendix 
A,
we find that the $\bar R$ dependence can be factored out to leave
\begin{eqnarray}
\frac{\pi}{2}\left(\frac{\sinh( \pi \szero )}{\cosh(\pi \szero )
-\cos\bigl(2\szero \ln(p/p_*)\bigr)}\right)
\bigl(\cot{\tilde\delta_3(p)}-i\bigr)
&=&\int_0^\Lambda\frac{q\sinh( \pi \szero )dq}
{\bigl[\cosh(\pi \szero )-\cos\bigl(2\szero \ln(q/p_*)\bigr)\bigr]
(p^2-q^2+i\epsilon)}\nonumber\\
&&\qquad\quad+\frac{\pi}{\szero }\sum_{|p_n|<\Lambda}
\frac{p_n^2}{p^2+p_n^2}-\frac{1}{\hat v_3(p/\Lambda,\Lambda)}.
\end{eqnarray}
To write this in a form which is explicitly independent of $\Lambda$, we
introduce the rescaled variable $\hat q=q/\Lambda$ and substitute in our
solution for $\hat v_3$ from  Eq.~(\ref{eq:solution}). This gives
\begin{equation}\label{eq:dwreprel}
\frac{\pi}{2}
\left(\frac{\sinh( \pi \szero )}{\cosh(\pi \szero )-\cos\bigl(2\szero
\ln(p/p_*)\bigr)}\right)
\bigl(\cot{\tilde\delta_3(p)}-i\bigr)
=\int_{-1}^1 d\hat q\,h(\hat q)-\int_C d\hat q\,h(\hat q)
-2\pi i\sum_{|p_n|<\Lambda}{\cal R}\left[h(\hat q),ip_n/\Lambda\right]
-\sum_{n=0}^\infty C_{2n}p^{2n},
\end{equation}
where $h(\hat q)$ is defined above in Eq.~(\ref{eq:jdef}) and we used the 
fact
that the sum over the bound states can be expressed in terms of the residues 
at
the bound-state poles (see Eq.~(\ref{eq:bsres})).

The two integrals over $h(\hat q)$ can be combined to form a single integral
round a closed contour. The $i\epsilon$ prescription in the first integral
implies that the contour of integration goes above the propagator
pole at $\hat q=-\hat p$ and below the one at $\hat q=\hat p$.
The contour thus encloses the pole at positive $\hat q$. It also encloses
the bound state poles at $\hat q=ip_n/\Lambda$ with $p_n<\Lambda$. Using
Cauchy's theorem, we find that the residues from these exactly cancel the
sum over bound states in Eq.~(\ref{eq:dwreprel}).

We are then left with a DW effective-range expansion,
\begin{equation}\label{eq:dwere}
\left(\frac{\sinh( \pi \szero )}{\cosh(\pi \szero 
)-\cos\bigl(2\eta(p)\bigr)}
\right)\bigl(\cot{\tilde\delta_3(p)}-i\bigr)=
\cot\left(\eta(p)+\frac{i\pi \szero }{2}\right)
-\frac{2}{\pi}\sum_{n=0}^\infty C_{2n}p^{2n},
\end{equation}
where $\eta(p)=-\szero \ln(p/p_*)$ is defined in Eq.~(\ref{eq:eta}). Make 
use
of trigonometric additon fomulae, one can show that the imaginary parts of
the left and right sides are equal. In this expansion, all nonanalytic
behaviour has been subtracted or factored out into the trigonometric 
functions
of $\ln p$. The remaining energy dependence can be expanded in powers of 
$p^2$
and the terms correspond directly to perturbations around the limit cycle.

The roles of the limit-cycle potential and its marginal perturbation
$C_0$ are not obvious from Eq.~(\ref{eq:dwere})
and so, to clarify these, we construct the total phase shift, $\delta$.
Making use of Eqs.~(\ref{eq:dwere}) and
(\ref{eq:smatrix0}), the corresponding full $S$-matrix can be written
\begin{equation}\label{eq:smatrix}
e^{2i\delta(p)}=i\frac{Z^*(p)}{Z(p)},
\end{equation}
where
\begin{equation}
\qquad
Z(p)=\cos\left(\eta(p)+\frac{i\pi \szero }{2}\right)
-\frac{2}{\pi}\,\sin\left(\eta(p)+\frac{i\pi \szero }{2}\right)
\sum_{n=0}^\infty C_{2n}p^{2n}.
\end{equation}

For the limit-cycle solution of Eq.~(\ref{eq:vs0}) with no perturbations
(all $C_{2n}=0$), we have
\begin{equation}
Z(p)=\cos\left(\eta(p)+\frac{i \pi \szero }{2}\right).
\end{equation}
Comparing this to Eq.~(\ref{eq:smatrix0}) for the pure long-range force,
we see that the phase $\eta(p)$ of the waves near the origin has been
shifted by $\pi/2$. This implies, for example, that the bound states lie at
$p=ie^{-\pi/2\szero }p_n$, which correspond to the geometric means of the 
bound
state energies for $v_3=0$. This potential thus corresponds to the system 
which
is ``furthest away" from the one with $v_3=0$, in the sense that it leads to
the maximum possible changes to physical observables.

To elucidate the role of the marginal perturbation, it is convenient to
express its coefficient in terms of an angle $\sigma$,
\begin{equation}
C_0=-\frac{\pi}{2}\cot\sigma,
\end{equation}
and to redefine the other short-distance parameters according to
\begin{equation}
\frac{2}{\pi}\sum_{n=1}^{\infty}C_{2n}p^{2n}
=\frac{\csc\sigma\sum_{n=1}^{\infty}C'_{2n}p^{2n}}
{\sin\sigma+\cos\sigma\sum_{n=1}^{\infty}C'_{2n}p^{2n}}.
\end{equation}
In terms of the new parameters, $\sigma$ and $C'_{2n}\,,n\geq1$ we find that
the $S$-matrix can still be written in the form of Eq.~(\ref{eq:smatrix}),
but with $Z(p)$ replaced by
\begin{equation}
Z^\prime(p)=\sin\left(\eta(p)+\sigma+\frac{i \pi \szero }{2}\right)
+\cos\left(\eta(p)+\sigma+\frac{i \pi \szero }{2}\right)
\sum_{n=1}^\infty C'_{2n}p^{2n}.\label{eq:bestZ}
\end{equation}

This form makes it clear that $\sigma$ (or rather $C_0$) just has the effect
of shifting the phase $\eta(p)$. Scattering observables depend only on the
combination
\begin{equation}
p'_*=e^{-\sigma/\szero }p_*,
\end{equation}
and not on $p_*$ and $C_0$
separately. This shows that $p_*$ and $C_0$ play the same role in 
determining
the phase of the wave functions for $R\rightarrow 0$ (the self-adjoint
extension of the Hamiltonian). We can use $C_0$ to change the self-adjoint
extension from the initial one specified by $p_*$ to any other.
In particular, $\sigma=0$ (corresponding to the trivial fixed point) leaves
the initial extension unchanged, whereas $\sigma=\pi/2$ ($C_0=0$)
produces the largest possible change in the extension, as already noted.
Furthermore, there is a one-to-one mapping between all possible limit-cycle
solutions (obtained by varying $C_0$ from $-\infty$ to $+\infty$ or
equivalently $\sigma$ between $0$ and $\pi$) and the self-adjoint
extensions. This is in marked contrast to the relationship between $p_*$
and the self-adjoint extensions, where infinitely many equivalent $p_*$'s
give the same extension.

The bound states of the system are given by the zeros of $Z'(p)$. For any
short-range potential, the bound states still accumulate at zero energy,
as expected since they are consequences of the inverse-square tail of the
long-range potential. The shallower bound states are insensitive
to the short-range perturbations, $C'_{2n}$ with $n\geq 1$. Their positions 
are
controlled by $p'_*$ rather than the original $p_*$.
If we choose $p_*$ to give the correct shallow Efimov states, we can set
$\sigma=0$ and use the expansion around the trivial fixed point. This 
corresponds
to expanding a DW $K$-matrix and has the form
\begin{equation}
\frac{\tan\tilde\delta_3(p)}{\sinh(\pi \szero )}
=\frac{\sum_{n=1}^\infty C'_{2n}p^{2n}}
{\cosh(\pi \szero )-\cos\bigl(2\eta(p)\bigr)
+\sin\bigl(2\eta(p)\bigr)\sum_{n=1}^\infty C'_{2n}p^{2n}}.
\end{equation}

The deeper bound states are of course strongly affected by the higher-order
perturbations and will not in general follow the simple constant-ratio 
pattern
of the Efimov states. At some point they will fall outside the range of
validity of the EFT and so we can say nothing about them, except that some
short-range physics must act to ensure the existence of a ground state.

\section{Finite Scattering Length}

In the general case of finite scattering length it no longer makes sense to 
expand
observables in terms of the hyperangular wavefunctions, instead we must 
consider a
much more general RG equation. In particular, the low-energy scales now
include $\gamma=1/a_2$ where $a_2$ is the two body scattering length. This 
can also
be thought of as the typical momentum in the low-lying bound or virtual 
state of
two particles. The coupling between waves with different hyperangular 
momenta
means that the states should now depend on the relative momentum $k$ of a
pair, as well as the total energy, expressed as a momentum $p$. Nonetheless
Efimov's separation of variables still applies whenever the distances
between the three particles are all much less than the two-body scattering 
length
\cite{efimov}. It therefore controls the short-distance behaviour of the
three-body wave function in an EFT with contact interactions and hence the 
scaling
of the short-range three-body interactions. For attractive three-body 
systems, the
the basic structure of the RG evolution will be similar to that in the 
previous
section since the wave functions at small $R$ will still be dominated by the
$\szero$ channel.

The truncated Green's function is given by Eq.~(\ref{eq:fullgreen}) 
restricted
to states inside the energy range $-\Lambda^2/M\le E\le\Lambda^2/M$:
\begin{equation}\label{eq:truncgreen}
G_2(p)=\frac{M}{4\pi}\sum_{|p_n|<\Lambda}\frac{|\Psi_n\rangle\langle 
\Psi_n|}{p^2+p_n^2}
+\frac{M}{2\pi^2}\int_{-\gamma^2}^{\Lambda^2}d(q^2)\frac{1}{p^2-q^2+i\epsilon}\left[
|\Psi_{q,i\gamma}\rangle\langle \Psi_{q,i\gamma}|+\vartheta(q^2)
\frac{2}{\pi}\int_0^q dk |\Psi_{q,k}\rangle\langle \Psi_{q,k}|\right].
\end{equation}
Inserting this into Eq.~(\ref{eq:potdiff}) and using Eq.~(\ref{eq:potform})
we obtain a differential equation for the potential
\begin{eqnarray}
\frac{\partial V_3(p,\gamma,k,k';\Lambda)}{\partial\Lambda}
&=&-\,\frac{M\bar R^4}{2\pi^2}\frac{1}{p^2-\Lambda^2}\Biggl[
V_3(p,\gamma,k,i\gamma;\Lambda)|\Psi_{\Lambda,i\gamma}(\bar R,\bar\Omega)|^2
V_3(p,\gamma,i\gamma,k';\Lambda)\nonumber\\
&&\qquad\qquad\qquad
+\frac{2}{\pi}\int_0^\Lambda dk'' 
V_3(p,\gamma,k,k'';\Lambda)|\Psi_{\Lambda,k''}
(\bar R,\bar\Omega)|^2
V_3(p,\gamma,k'',k';\Lambda)\Biggr]\nonumber\\
&&-\frac{M\bar R^4}
{4\pi}\sum_{n=0}^\infty\frac{|\Psi_n(\bar R,\bar\Omega)|^2}{p^2+p_n^2}\,
\delta(\Lambda-p_n)V_3(p,\gamma,k,ip_n/3;\Lambda)
V_3(p,\gamma,ip_n/3,k';\Lambda).
\end{eqnarray}

For $R\ll a_2$ all the DW's will be dominated by the $\szero$ channel and 
will
therefore tend to the forms
\begin{eqnarray}
|\Psi_{p,k}(R,\Omega)|^2&\sim& \frac1p{\cal D}_3(p,k,\gamma,p_*)\,
|\xi(\Omega)|^2\frac{\sin^2\left(\szero 
\ln(p_*R)-\theta\right)}{R^4},\nonumber\\
|\Psi_{p,i\gamma}(R,\Omega)|^2&\sim& {\cal D}_2(p,\gamma,p_*)\,
|\xi(\Omega)|^2\frac{\sin^2\left(\szero 
\ln(p_*R)-\theta\right)}{R^4},\nonumber\\
|\Psi_{n}(R,\Omega)|^2&\sim& p_n^2\,\Dbn(\gamma,p_*)\,
|\xi(\Omega)|^2\frac{\sin^2\left(\szero \ln(p_*R)-\theta\right)}{R^4}.
\label{eq:generaldws}
\end{eqnarray}
The various normalisation functions, ${\cal D}_3$, ${\cal D}_2$ and $\Dbn$,
are determined by the external boundary conditions on the DW's. They can be
found by solving the full Faddeev equations (or equivalents). These 
functions are
essential to the RG discussion, since it is through them that information 
about
the long-distance physics is communicated to short distances.

The forms of the DW's above were chosen to ensure that the normalisation 
functions
are dimensionless. They have a common dependence on the hyperangles given by 
the
function
\begin{equation}
\xi(\Omega)={\cal A}_{s_0}\sum_{i=1}^3\frac{2}{\sin(2\varphi_i)}
\sinh\left(\frac{s_0\pi}{2}-s_0\varphi_i\right),
\end{equation}
where $\varphi_i$ is the hyperangle defined in appendix A.
Since this dependence can be factored out of the Green's function $G_2(p)$ 
at
small $R$, it is unimportant to the RG discussion. In this context, it is
convenient to define a Green's function with this common short-distance
behaviour divided out,
\begin{equation}
{\cal G}_2(p,\gamma,p_*)=\frac{2\pi^2\bar R^4}{M}\frac{
G_2(p; \bar R,\bar\Omega;\bar R,\bar\Omega)}{
|\xi(\bar\Omega)|^2\sin^2\left(\szero \ln(p_*\bar R)-\theta\right)}.
\label{eq:g2short}
\end{equation}

Using the forms of the DW's given in Eq.~(\ref{eq:generaldws}) we define our
rescaled potential by
\begin{equation}
\hat V_3(\hat p,\hat\gamma,\hat k,\hat 
k';\Lambda)=\frac{M}{2\pi^2}|\xi(\bar\Omega)|^2
\sin^2\bigl(\szero \ln(p_*\bar R)-\theta\bigr)V_3(p,\gamma,k,k';\Lambda),
\end{equation}
where, as usual, $\hat p=p/\Lambda$, etc.
We also require rescaled versions of the normalisation functions $\cal D$,
which we define by
\begin{equation}
\hat{\cal D}_3(\hat k,\hat\gamma,\Lambda)=
{\cal D}_3(\Lambda,\Lambda\hat k,\Lambda\hat\gamma,p_*),\qquad
\hat{\cal D}_2(\hat\gamma,\Lambda)=
{\cal D}_2(\Lambda,\Lambda\hat\gamma,p_*),\qquad
\hDbn(\hat\gamma,\Lambda)=
\Dbn(\Lambda\hat\gamma,p_*),\label{eq:defineDWhat}
\end{equation}
Since $p_*$ only occurs in these dimensionless functions as a ratio with
$\Lambda$ (the only other scale) we have suppressed the dependence on it
to  simplify notation. We also define a similarly rescaled Green's function,
\begin{equation}
\hat{\cal G}_2(\hat\gamma,\Lambda)={\cal 
G}_2(\Lambda,\Lambda\hat\gamma,p_*).
\label{eq:hg2}
\end{equation}

The rescaled potential $\hat V_3$ satisfies the RG equation
\begin{eqnarray}
\Lambda\frac{\partial \hat V_3}{\partial\Lambda}
&=&\hat p\frac{\partial \hat V_3}{\partial\hat p}
+\hat\gamma\frac{\partial \hat V_3}{\partial\hat\gamma}
+\hat k\frac{\partial \hat V_3}{\partial\hat k}+\hat k'\frac{\partial \hat 
V_3}
{\partial\hat k'}\nonumber\\
&&+\frac{1}{1-\hat p^2}\Biggl[
\hat V_3(\hat p,\hat\gamma,\hat k,i\hat\gamma;\Lambda)\hat{\cal D}_2
(\hat\gamma,\Lambda)
\hat V_3(\hat p,\hat\gamma,i\hat\gamma,\hat k';\Lambda)\nonumber\\
&&\qquad\qquad
+\frac{2}{\pi}\int_0^1 d\hat k'' \hat V_3(\hat p,\hat\gamma,\hat k,\hat 
k'';\Lambda)
\hat{\cal D}_3(\hat k'',\hat\gamma,\Lambda)
\hat V_3(\hat p,\hat\gamma,\hat k'',\hat k';\Lambda)\Biggr]\nonumber\\
&&+\frac{\pi}{2}\sum_{n=0}^\infty\hat{\cal D}_B^{(n)}(\hat\gamma,\Lambda)
\frac{1}{1+\hat p^2}\delta(\hat p_n(\Lambda)-1)
\hat V_3(\hat p,\hat\gamma,\hat k,i\hat p_n/3;\Lambda)
\hat V_3(\hat p,\hat\gamma,i\hat p_n/3,\hat k';\Lambda).
\end{eqnarray}
The boundary conditions on $\hat V_3$ are that it be analytic in $\hat p^2$,
$\hat\gamma$, $\hat k^2$ and $\hat k'^2$. Again these follow from our 
requirement
that the potential arises from six-point contact terms in the effective 
Lagrangian.

Note that in the $\hat{\cal D}$'s and the rescaled bound-state momenta
$\hat p_n(\Lambda)$ the only explicit dependence on $\Lambda$ occurs in the 
ratio
$\Lambda/p_*$. (The scales $p_*$ and $\Lambda$ are the only ones in these
dimensionless functions.) The scale $p_*$ is introduced, as before, by the
self-adjoint extension which fixes the phase of the waves for $R\rightarrow 
0$.
All quantities appearing in the RG equation are thus invariant under the
transformation of $p_*$ in Eq.~(\ref{eq:p*transf}) and hence are periodic in 
$p_*$.
Furthermore, since $p_*$ and $\Lambda$ always appear as a ratio in the RG 
equation,
that equation must be unchanged under
\begin{equation}\label{eq:transf3}
\Lambda\rightarrow \Lambda e^{-n\pi/\szero },\qquad n\in\mathbb Z.
\end{equation}
In interpreting this for the bound state momenta $\hat p_n(\Lambda)$ we need 
to
be careful. Depending on how we label the states, these momenta may not obey
the symmetry individually. However the complete set must always obey it
collectively.

To see how this happens, consider the entire rescaled spectrum, $\hat p_n$, 
at
some initial scale $\Lambda=\Lambda_1$. As $\Lambda$ decreases the rescaled
bound state momenta increase, until at $\Lambda=\Lambda_1 e^{-\pi/\szero }$ 
the
spectrum has shifted so that $\hat p_n(\Lambda)=\hat p_{n+1}(\Lambda')$.
This is similar to what happened in the case of infinite scattering length,
except that a new bound state must have appeared at threshold
and moved down to $\hat p_0(\Lambda_1)$. This is possible because in 
following
the RG flow we keep the rescaled quantity $\hat\gamma$ fixed, not the 
physical
value $\gamma$. The spectrum at $\Lambda=\Lambda_1 e^{-\pi/\szero }$ is thus
identical to that at $\Lambda_1$. If we label the bound states so that
$\hat p_0(\Lambda)$ always refers to the shallowest, then each of the
$\hat p_n(\Lambda)$ will be invariant under Eq.~(\ref{eq:transf3}).

The general RG equation is complicated because it includes couplings between
the different three-body channels. As a starting point for solving this
equation, we consider a solution which is independent of the two relative 
momenta,
$\hat k$ and $\hat k'$. This simplifies the equation considerably since it 
allows
us to divide through by $\hat V_3(\hat p,\hat\gamma;\Lambda)^2$ to obtain a 
linear
equation for $1/\hat V_3$,
\begin{eqnarray}
&&\displaystyle{
\Lambda\frac{\partial}{\partial\Lambda}\left(\frac{1}{\hat V_3}\right)=
\hat p\frac{\partial}{\partial\hat p}\left(\frac{1}{\hat V_3}\right)+
\hat\gamma\frac{\partial}{\partial\hat\gamma}\left(\frac{1}{\hat 
V_3}\right)-
\frac{1}{1-\hat p^2}\hat P(\hat\gamma,\Lambda)
+\frac{\pi}{2}\sum_{n=0}^\infty\hat{\cal D}_B^{(n)}(\hat\gamma,\Lambda)
\frac{1}{1+\hat p^2}\delta(\hat p_n(\Lambda)-1),}\label{eq:3bdwrg2}
\end{eqnarray}
where we have introduced the function
\begin{equation}
\hat P(\hat\gamma,\Lambda)=
\hat {\cal D}_2(\hat\gamma,\Lambda)
+\frac{2}{\pi}\int_0^1 d\hat k''\hat{\cal D}_3(\hat 
k'',\hat\gamma,\Lambda).\label{eq:proj}
\end{equation}
This function can be thought of as the (rescaled) short-distance part of the
projection operator onto continuum states with energy $E=\Lambda^2/M$.

Our strategy for solving Eq.~(\ref{eq:3bdwrg2}) will mirror that used in the
simpler case of infinite scattering length. As before, we can immediately 
write
down a solution to the continuous equation using an equivalent to the basic 
loop
integral, Eq.~(\ref{eq:defineI1}). Again this will not satisfy the 
analyticity
boundary conditions because of the behaviour of the integrand at the 
endpoint
$\hat q=0$. Furthermore this solution will not generate the discontinuities 
at the
bound states. We must therefore find a generalisation of 
Eq.~(\ref{eq:splitI})
and write the solution as a contour integral that avoids the point $\hat 
q=0$.

\begin{figure}
\begin{center}
\includegraphics[width=10cm,angle=0]{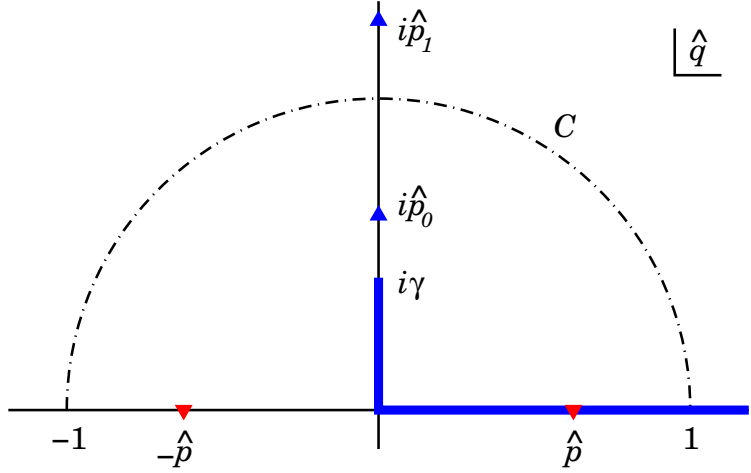}
\caption{The contour $C$ used to construct the DWRG solution
$\hat J(\hat p, \hat\gamma,\Lambda)$.
The bound-state poles of the integrand $H(\hat q)$ occur at $\hat q=i\hat 
p_n$.
The branch cut running down the imaginary axis corresponds to the $2+1$ 
elastic
continuum. This is joined by a cut starting at $\hat q=0$ corresponding to 
the
three-body continuum. There are also propagator poles at $\hat q=\pm\hat 
p$.}
\label{fig:jpoles2}
\end{center}
\end{figure}

Guided by the results for infinite scattering length, we define an
integral along the same contour $C$ as before,
\begin{equation}
\hat J(\hat p,\hat\gamma,\Lambda)=\int_C d\hat q\,H(\hat q),\label{eq:J3b}
\end{equation}
where
\begin{equation}
H(\hat q)=\frac{i}{\pi}\frac{\hat q}{\hat p^2-\hat q^2}\,
\hat{\cal G}_2(\hat\gamma/\hat q,\hat q\Lambda).
\label{eq:H3b}
\end{equation}
The contour $C$ is shown in Fig.~\ref{fig:jpoles2} along with the 
singularity
structure of $H$. This structure follows straightforwardly from the form of
$\hat{\cal G}_2$ and is derived in appendix B.

In appendix B we also show that the function $\hat P(\hat\gamma,\Lambda)$ in
Eq.~(\ref{eq:3bdwrg2}) can be rewritten in terms of $\hat{\cal G}_2$ as
\begin{equation}
\hat P(\hat\gamma,\Lambda)
=\frac{i}{\pi}\Bigl(\hat{\cal G}_2(\hat\gamma,\Lambda)
-\hat{\cal G}_2(-\hat\gamma,-\Lambda)\Bigr).\label{eq:analyticP}
\end{equation}
This result can then be used to show that $\hat J(\hat 
p,\hat\gamma,\Lambda)$
satisfies the continuous part of Eq.~(\ref{eq:3bdwrg2}). In addition, the 
bound
state poles in $H$ cross the contour of integration at $\hat p_n=1$ and so
produce the necessary discontinuities.
Finally, the contour never approaches the branch points at $\hat q=0$ and
$\hat q=i\hat\gamma$ and so the integral is analytic in the small scales.
This allows us to take our nontrivial solution to the DWRG equation to be
\begin{equation}
\hat V_3^{(0)}(\hat p,\hat\gamma,\Lambda)=\bigl[\hat J(\hat p,\hat\gamma,
\Lambda)\bigr]^{-1}.
\end{equation}

The periodic dependence on $\Lambda$ of all quantities appearing the DWRG
equation (\ref{eq:3bdwrg2}) implies that our solution
$\hat V_3^{(0)}(\hat p,\hat\gamma,\Lambda)$ must also be periodic. Like
the analogous solution for infinite scattering length, it is therefore a
limit-cycle of the DWRG.

As a result, it must have an infinite number of discontinuities as
$\Lambda\rightarrow0$, which follow a periodic pattern according to this 
symmetry.
At first glance this may seem odd since these discontinuities are
associated with the truncation of bound states and there are only a finite 
number
of bound states in the truncated Green's function. However as $\Lambda$ is
lowered for fixed $\hat\gamma$, new bound states appear in the rescaled 
spectrum
as discussed above

Now that we have obtained a limit-cycle solution, the analysis proceeds 
along
very similar lines to that in the previous section. The perturbations in 
$\hat p$
and $\hat\gamma$ are simple to derive and, since Eq.~(\ref{eq:3bdwrg2}) is 
linear,
they give exact solutions to the DWRG equation:
\begin{equation}
\frac{1}{\hat V_3(\hat p,\hat\gamma,\Lambda)}=\frac{1}{\hat V_3^{(0)}
(\hat p,\hat\gamma,\Lambda)}+\sum_{n,m=0}^\infty C_{2n,m}\Lambda^{2n+m}
\hat p^{2n}\hat\gamma^m.\label{eq:general3b}
\end{equation}
The main features of the power-counting are unaffected by the addition of 
the
new scale $\gamma$. The LO perturbation is still marginal, and leads to a 
family
of limit-cycles parametrised by $C_{0,0}$. Although it is more difficult to 
show,
the analysis of the infinite-scattering-length case suggests that the 
marginal
perturbation $C_{0,0}$ and the scale $p_*$ must represent the same 
three-body physics
in the EFT.

The previous power-counting is
supplemented by terms proportional to powers of $\gamma$. In many systems, 
such
as those of interest in nuclear physics, the two-body scattering lengths and 
hence
$\gamma$ are fixed. This means that these terms can not be determined 
separately
but must be absorbed into the coefficients of the energy-dependent 
perturbations.
However in atomic systems with tunable Feshbach resonances, it may be 
possible to
vary $\gamma$ and hence to disentangle the energy- and $\gamma$-dependent
perturbations.

The full set of perturbations also includes ones that depend on the relative
momenta. The forms of these have been derived and their eigenvalues show 
that
they occur at the same orders as the corresponding energy-dependent 
perturbations.
The full expressions are very unwieldy and so the interested reader is 
referred to
Ref.~\cite{tbthesis}.

Having obtained a solution to the RG equation we can substitute it back into 
the
LS equation to obtain the corresponding DW $T$-matrix. The algebra again 
follows
that for systems with infinite scattering length. In the present case, the
truncated Green's function involves an integral around the branch cut 
corresponding
to elastic scattering of a single particle on a bound pair below the 
three-body
threshold. This $2+1$ continuum cut runs along the imaginary $\hat q$-axis 
from
$i\hat\gamma$, as shown in Fig.~\ref{fig:jpoles2}. Using the form for the
three-body force given in Eq.~(\ref{eq:general3b}) we find that the DW 
$T$-matrix
element for $2+1$ elastic scattering can be written
\begin{equation}
\frac{2\pi^2}{M}\,\frac{{\cal D}_2(p,\gamma,p_*)e^{2i\delta_2(p)}}
{\langle\Psi^-_{p,i\gamma}|\tilde T_3(p)|\Psi^+_{p,i\gamma}\rangle}
=-{\cal G}_2(p,\gamma,p_*)+\sum_{n,m=0}^\infty C_{2n,m}
p^{2n}\gamma^m.\label{eq:3beff}
\end{equation}
Using the properties of ${\cal G}_2$ in Appendix B, one can show that
$\tilde T_3(p)$ respects unitarity for $p^2<0$, as it should. For $p^2>0$,
the coupling to three-body breakup channels, mean that $T$-matrix for $2+1$
scattering does not repect unitarity on its own. The same effective 
potential
could be used to calculate amplitudes for breakup and $3\rightarrow3$ 
scattering.
However, in practice this is likely to require general momentum-dependent
potentials to describe all possible interaction channels.

The DW $T$-matrix can be related to the elastic-scattering phase shift,
$\delta$, by
\begin{equation}
\frac{e^{2i\delta_2(p)}}{\langle\Psi^-_{p,i\gamma}|\tilde T_3(p)
|\Psi^+_{p,i\gamma}\rangle}
=-\frac{M}{4\pi} \bigl[\cot\bigl(\delta(p)-\delta_2(p)\bigr)-i\bigr],
\end{equation}
where $\delta_2$ is the phase shift produced by pairwise forces only. This
allows us to rewrite Eq.~(\ref{eq:3beff}) in the form of a DW
effective-range expansion,
\begin{equation}
\frac{\pi}{2}\,{\cal D}_2(p,\gamma,p_*)
\bigl[\cot\bigl(\delta(p)-\delta_2(p)\bigr)-i\bigr]-{\cal G}_2(p,\gamma,p_*)
=-\sum_{n,m=0}^\infty C_{2n,m}p^{2n}\gamma^m.
\label{eq:elasticeff}
\end{equation}

\section{Discussion}

In this paper we have applied a DW version of the RG to three-body 
scattering.
This is based on cutting off the full solutions in the presence of 
long-range
interactions \cite{bb1}. In three-body systems, these long-range forces are
generated by point-like interactions between pairs of particles. By imposing
the cut-off on the DW's we ensure that it affects only the contributions of
the short-range three-body forces, and does not alter the two-body physics.
Demanding that observables are independent of the cut-off then leads to an 
RG
equation for the renormalised three-body interactions.

As shown by Efimov \cite{efimov}, a three-body system  with contact 
interactions
can be described by an ISP in the region where they are all close together. 
Hence
the RG in presence of this rather singular potential controls the behaviour 
of the
three-body forces and determines their power counting.

For systems of three bosons or two neutrons and a proton with total spin
$\frac{1}{2}$, the resulting ISP is attractive. This leads to quite 
different
behaviour compared with previous examples studied with this type of RG
\cite{bmr,bb1}. The wave functions show oscillatory behaviour at short 
distances
and a self-adjoint extension is needed to fix their phase and hence give
well-defined eigenfunctions \cite{meetz,case,pp,bc}. These oscillatory 
functions
control the RG flow of the short-distance potential, which tends to a limit 
cycle
\cite{wilslc} rather than a nontrivial fixed point.

These results are general for an attractive ISP in any number of dimensions.
The lack of a ground state for such a potential means that it has deeply
bound states that lie outside the domain of validity of our effective 
theory.
This means that the cut-off should truncate not only high-energy scattering 
states
but also these deeply bound states. This leads to a three-body force that 
has
periodic discontinuities, but is single-valued.

The resulting power counting for perturbations around the limit cycle starts
with a marginal term. From its effect on observables, we have seen that this
term is equivalent to changing the parameter in the self-adjoint extension
or, in other words, the phase of the short-distance wave functions. To 
express
the resulting power counting in a form which matches that usually used
\cite{wein79,wein}, we can assign an order $d=\mu-1$ to a term in the 
rescaled
potential proportional to $\Lambda^\mu$. Then a term proportional to 
$p^{2n}$
(or $E^n$) appears at order $d=2n-1$. The coefficents in our potential have
direct, and relatively simple, connection to scattering observables.

The same behaviour is found in an attractive three-body system with finite 
two-body
scattering length, $a_2$, since the wave functions tend to the Efimov form 
at short
distances ($R\ll a_2$). Hence the RG flow has a very similar pattern, 
tending to a
limit cycle. In this case the power counting should be extended to include 
terms
involving powers of the low-energy scale $\gamma=1/a_2$. A term proportional 
to
$p^{2n}\gamma^m$ is of order $d=2n+m-1$. For the energy-dependent 
perturbations,
this counting agrees with that found by Bedaque {\it et al.}~\cite{bghr} 
from the
STM equation. Because of the clean separation of the short- and long-range
physics in the DWRG approach we are able to state the counting in an 
algebraically
much simpler way.

We are also able to extend the counting to terms involving powers of 
$\gamma$.
This may not be relevant to applications in nuclear physics, where the 
two-body
scattering lengths are fixed, but in atomic systems it may be possible to 
use
Feshbach resonances to vary the scattering length, and hence to disentangle 
the two
types of perturbation.

The leading, marginal three-body force, or equivalently the choice of 
self-adjoint
extension, is of interest since it shows that one piece of three-body 
physics is
needed to determine low-energy three-body observables. As noted by Bedaque
{\it et al.}~\cite{bhvk}, this provides a way to understand the old 
observation of
the ``Phillips line,'' a correlation between the $J=\frac{1}{2}$ $nd$ 
scattering
lengths  and triton binding energies predicted by model nucleon-nucleon 
potentials
\cite{phillips}.

At leading order ($d=-1$), three-body observables are determined by the 
two-body
scattering length and the marginal three-body force. The first correction to 
this
arises from the two-body effective range, which appears at order $d=0$. 
Including
this has been shown to give good agreement with with $J=\frac{1}{2}$ $nd$
observables, in calculations using the STM equation \cite{bghr} and an 
equivalent
equation in coordinate-space \cite{tbthesis}. The first energy-dependent 
three-body
force appears only at next-to-next-to-leading order, $d=1$ in this counting.

It should be possible to extend our approach to three-body scattering above 
the
breakup threshold. However this will introduce couplings between various 
channels
and so will involve momentum-dependent perturbations in the effective 
potential.
This is likely to require similar analyses to those which have been applied 
to
simpler coupled-channel systems, such as that considered in 
Ref.~\cite{cgvk}.

\section*{Acknowledgments}

This work was supported by the EPSRC. MCB acknowledges the Institute for
Nuclear Theory, Seattle for hospitality during the programme on Effective
Field Theories and Effective Interactions where the seeds of these ideas 
were
sown, and T. Cohen, S. Coon, J. McGovern and R. Perry for useful 
discussions.
He is also grateful for discussions with A. C. Phillips, a much missed
colleague and friend.

\begin{appendix}

\section{Efimov wave functions}

We summarise here the essential elements of Efimov's approach to the 
three-body
problem \cite{efimov}, since this provides the wave functions we need to 
elucidate
the scaling of the the three-body interactions.

We consider the case of $s$-wave scattering. In general, the wavefunctions
$|\Psi_{p,k}\rangle$ must be found using the Faddeev equations or equivalent
(see, for example, Ref.~\cite{m3hp}). In the Faddeev formalism, the 
wavefunction
is broken into three components according to the pair of particles that 
interacted
last. The component in which particles $2$ and $3$ interacted last is 
denoted
$\psi^{(1)}_{p,k}(r_{23},r_1)$, where $r_{23}$ and $r_1$ are the usual 
Jacobian
coordinates for three bodies with equal masses.

For particles interacting via pairwise contact interactions, Efimov observed
that $\psi^{(i)}_{p,k}(r_{jk},r_i)$ satisfies a free two-dimensional 
Schr\"odinger
equation, subject to a boundary condition relating the wave function at the 
points
$r_{jk}=0$ and $r_i=r_{jk}/2$ where pairs of particles interact. This 
boundary
condition takes the form
\begin{equation}
\left[\frac{\partial \psi^{(i)}_{p,k}(R,\varphi_i)}{\partial\varphi_i}
\right]_{\varphi_i=0}
+\frac{8\lambda}{\sqrt{3}}
\psi^{(i)}_{p,k}(R,\pi/3)=\frac{R}{a_2}\psi_{p,k}^{(i)}(R,0),\label{eq:efimbc}
\end{equation}
where $\lambda$ is a factor related to the wave-function symmetry, and
$R$ and $\varphi_i$ are the hyperradius and hyperangle respectively,
and are defined by
\begin{equation}
R=\sqrt{\frac43r_i^2+r_{jk}^2},\qquad 
\varphi_i=\arctan{\frac{\sqrt{3}r_{jk}}{2r_i}}.
\end{equation}
For three bosons, the full wavefunction can be expressed as
\begin{equation}\label{eq:generalDW}
\Psi_{p,k}(R,\Omega)=\sum_{i=1}^3\frac{2}{R^2\sin(2\varphi_i)}
\psi^{(i)}_{p,k}(R,\varphi_i),
\end{equation}
where $\Omega$ represents five general hyperspherical coordinates that 
complement
$R$. (The other coordinates could be $\varphi_i$ and two angles each for
${\bf r}_i$ and ${\bf r}_{jk}$, but their exact specification will not be 
needed
here.)

We shall assume that the DW's are normalised by
\begin{eqnarray}
\int_0^\infty R^5dR\int d\Omega\,\Psi_{p,k}(R,\Omega)\Psi_{p',k'}(R,\Omega)
&=&\frac{\pi^2}{4}\delta(p^2-p'^2)\delta(k-k'),\\
\int_0^\infty R^5dR\int d\Omega\,\Psi_{p,i\gamma}(R,\Omega)
\Psi_{p',i\gamma}(R,\Omega)&=&\frac{\pi}{2}\delta(p^2-p'^2),\\
\int_0^\infty R^5dR\int d\Omega\,\Psi_n(R,\Omega)\Psi_n(R,\Omega)&=&1.
\end{eqnarray}

Because the boundary condition couples $R$ and $\varphi_i$, the equations 
are,
in general, extremely difficult to solve. However when $R\ll a_2$ the 
boundary
condition separates. In this limit we may label the states in terms of the
centre-of-mass energy and the hyperangular momentum, $s$. We can write the
solutions in the separable form
\begin{equation}
\psi_{p,s}^{(i)}(R,\varphi_i)={\cal 
A}_s\sin\left(\frac{s\pi}{2}-s\varphi_i\right)
u_{s}(p,R).
\end{equation}
This satisfies the angular Schr\"odinger equation subject to the boundary
condition of vanishing amplitude at $\varphi_i=\pi/2$ ($r_i=0$). The 
boundary
condition, Eq.~(\ref{eq:efimbc}), now results in a transcendental equation 
for $s$,
\begin{equation}\label{eq:danilov}
s\cos{\frac{s\pi}{2}}=\frac{8\lambda}{\sqrt{3}}\sin\frac{s\pi}{6}.
\end{equation}
This equation was also obtained much earlier by Danilov \cite{danilov}
using the momentum-space equation derived by Skorniakov and
Ter-Martirosian \cite{stm}.

The resulting radial Schr\"odinger equation has the form
\begin{equation}\label{eq:2dse}
-{\frac{d^2u_{s}(p,R)}{dR^2}}-\frac{1}{R}{\frac{du_s(p,R)}{dR}}
+\frac{s^2}{R^2}u_s(p,R)=p^2u_s(p,R),
\end{equation}
and so contains an ISP whose strength is determined by $s$, and hence 
depends on
the symmetry parameter. For three bosons or a neutron and deuteron with 
spin-$1/2$,
the symmetry parameter is $\lambda=1$, while for three nucleons with
spin-$3/2$ it is $\lambda=-1/2$. In the latter case, Eq.~(\ref{eq:danilov})
has real solutions and inverse-square potential provides a repulsive 
``centrifugal
barrier". The power counting for short-range interactions in the presence of 
this
potential can be derived using the RG method of Ref.~\cite{bb1}.

The case with $\lambda=1$ is more interesting, since the lowest solutions to
Eq.~(\ref{eq:danilov}) are imaginary,
\begin{equation}
s=\pm i\szero =\pm i1.006\cdots,
\end{equation}
and so they correspond to an attractive ISP. For small $R$ the solutions to 
the
corresponding radial equation have the forms
\begin{equation}
u_{\szero}(p,R)\propto R^{\pm i\szero }.
\end{equation}
These are both equally (ir)regular as $R\rightarrow 0$. They can also 
describe
flux disappearing or being created at the origin, corresponding to the 
classical
``fall into the centre'' which is possible for this potential \cite{newton}.
To obtain well-defined wave functions which respect flux conservation, we 
need
to impose a boundary condition on the wave functions for $R\rightarrow 0$,
requiring equal admixtures of the two complex solutions and fixing their 
relative
phase. Mathematically, this is known as choosing a self-adjoint extension of 
the
Hamiltonian in Eq.~(\ref{eq:2dse}), as discussed by Bawin and Coon \cite{bc}
and references therein.

The full solutions to the Schr\"odinger equation with an attractive ISP can 
be
written in terms of Bessel functions of imaginary order,
\begin{equation}\label{eq:isqdw}
u_{\szero}(p,R)=\sqrt\frac{p\pi}{2}\frac{1}{2i|\sin(\eta(p)+i\pi \szero 
/2)|}
\left[e^{i\eta(p)}J_{i\szero }(pR)-e^{-i\eta(p)}J_{-i\szero }(pR)\right],
\end{equation}
where $p=\sqrt{ME}$ and the normalisation has been fixed by requiring that
$u_{\szero}(p,R)\sim\sin(pR+\delta)/\sqrt{R}$ for large $R$. For small $R$, 
these
wave functions have the form
\begin{equation}\label{eq:isqdwr0}
u_{\szero}(p,R)\sim \sqrt{\frac{p\sinh(\pi \szero )}{\szero
\bigl[\cosh(\pi \szero )-\cos\bigl(2\eta(p)\bigr)\bigr]}}
\sin\bigl(\szero \ln(pR)+\eta(p)-\theta\bigr),
\end{equation}
where
\begin{equation}
\theta=\arg\Gamma(1+i\szero )+\szero \ln{2}.
\label{eq:theta}
\end{equation}
Demanding that these waves tend to a common, energy-independent form as
$R\rightarrow 0$ implies that $\eta(p)$ must be of the form
\begin{equation}\label{eq:eta}
\eta(p)=-\szero \ln(p/p_*),
\end{equation}
where $p_*$ is a parameter which fixes the phase of the sine-log 
oscillations (or
equivalently specifies the self-adjoint extension).

The physical meaning of $p_*$ can be seen by noting that the $S$-matrix for 
this
system is
\begin{equation}\label{eq:smatrix0}
e^{2i\delta_2(p)}=i\frac{\sin(\eta(p)-i\pi \szero /2)}{\sin(\eta(p)+i\pi 
\szero /2)}.
\end{equation}
This has a pole at $p=ip_*$, implying that $E=-p_*^2/M$ corresponds to a
bound state. In fact, as shown by Efimov, there is an infinite tower of
bound states with energies $E=-p_n^2/M$ where
\begin{equation}\label{eq:bsmom}
p_n=p_*e^{n\pi/\szero }.
\end{equation}

The bound states accumulate at zero energy and extend downwards in a 
geometric
pattern with no ground state. The wave functions of the bound states are
\begin{equation}
u_{\szero}^{(n)}(R)=\sqrt{\frac{2\sinh(\pi \szero )}{\pi \szero }}\,
p_nK_{i\szero }(p_nR),
\end{equation}
where $K_m(x)$ denotes a modified Bessel function of the third kind.
Near the origin these have the form
\begin{equation}\label{eq:isqbsr0}
u_{\szero}^{(n)}(R)\sim\frac{\sqrt{2}}{\szero } p_n(-1)^{n+1}
\sin\left(\szero \ln{p_*R}-\theta\right).
\end{equation}

\section{Rescaled projection operator}

The short-distance Green's function defined in Eq.~(\ref{eq:g2short}) has 
the
spectral representation
\begin{equation}
{\cal G}_2(p,\gamma,p_*)=\int_{-\gamma^2}^\infty d(q^2)\frac{1}
{p^2-q^2+i\epsilon}\left[{\cal D}_2(q,\gamma,p_*)
+\vartheta(q^2)\frac{2}{\pi}\int_0^q\frac{dk}{q}{\cal D}_3(q,k,\gamma,p_*)
\right]+\frac{\pi}{2}\sum_{n=0}^\infty{\cal D}_B^{(n)}(\gamma,p_*)
\frac{p_n^2}{p^2+p_n^2}.\label{eq:defineDG}
\end{equation}
Using the result,
\begin{equation}
\lim_{\epsilon\rightarrow0}\int_{0}^\infty \frac{f(x)dx}{x_0-x\pm 
i\epsilon}=
{\cal P}\int_0^\infty \frac{f(x)dx}{x_0-x}\mp i\pi f(x_0),
\end{equation}
where ${\cal P}$ denotes the principal value, we find that, for real $p$,
\begin{equation}\label{eq:ImDg}
\frac{i}{\pi}\left[{\cal G}_2(p,\gamma,p_*)-{\cal G}_2(-p,\gamma,p_*)\right]
={\cal D}_2(p,\gamma,p_*)+\frac{2}{\pi}\int_0^p\frac{dk}{p}{\cal 
D}_3(p,k,\gamma,p_*).
\end{equation}
In this equation ${\cal G}_2(-p,\gamma,p_*)$ is found by analytically 
continuing $p$
through the upper half of the complex $p$ plane to $-p$. Since, by its 
definition,
${\cal G}_2(p,\gamma,p_*)$ is real for pure imaginary $p$, its value at
negative real $p$ is the complex conjugate of that at positive real $p$ and 
hence
${\cal G}_2(-p,\gamma,p_*)$ corresponds to a $-i\epsilon$ prescription at 
the
propagator pole.

By rescaling Eq.~(\ref{eq:ImDg}), we obtain an expression for the rescaled
projection operator defined in Eq.~(\ref{eq:proj}):
\begin{equation}
\hat P(\hat\gamma,\Lambda)
=\frac{i}{\pi}\Bigl(\hat{\cal G}_2(\hat\gamma,\Lambda)
-\hat{\cal G}_2(-\hat\gamma,-\Lambda)\Bigr),
\end{equation}
where $\hat{\cal G}_2$ is the rescaled Green's function defined in
Eq.~(\ref{eq:hg2}).

That rescaled Green's function has the representation
\begin{eqnarray}
\hat{\cal G}_2(\hat\gamma,\Lambda)&=&\int_{-\hat\gamma^2}^\infty d(\hat 
q^2)\,
\frac{1}{1-\hat q^2+i\epsilon}\left[\hat{\cal D}_2\left(
\frac{\hat\gamma}{\hat q},\hat q\Lambda\right)
+\vartheta(\hat q^2)\frac{2}{\pi}\int_0^1d\hat k\,\hat {\cal 
D}_3\left(\frac{\hat k}{\hat q}
,\frac{\hat\gamma}{\hat q},\hat q\Lambda\right)\right]\nonumber\\
&&+\frac{\pi}{2}\sum_{n=0}^\infty\frac{\hat{\cal 
D}_B^{(n)}(\hat\gamma,\Lambda)
\hat p_n(\Lambda)^2}{1+\hat p_n(\Lambda)^2}.\label{eq:defDGhat}
\end{eqnarray}
The analytic properties of this function are needed since it forms part of
the integrand $H(\hat q)$ of Eq.~(\ref{eq:H3b}), which is used to
construct the RG limit-cycle solution. From its spectral representation, we 
see
that $\hat{\cal G}_2(\hat\gamma,\Lambda)$ has poles at the bound state
momenta, $\hat q=i\hat p_n(\Lambda)$. The integral term results
in a branch cut down the imaginary axis from $\hat q=i\hat\gamma$ and then
along the positive real axis.

\end{appendix}

\end{document}